# Balanced Area Deprivation Index (bADI): Enhancing social determinants of health indices to strengthen their association with healthcare clinical outcomes, utilization and costs

Mohammad Amin Morid[1,2], PhD, Robert E Tillman[1], PhD, Eran Halperin, PhD[1,3]

[1] Optum AI Labs, UnitedHealth Group, Eden Prairie, MN, USA.

[2] Department of Information Systems and Analytics, Leavey School of Business, Santa Clara University, CA, USA.

[3] Department of Computer Science, University of California, Los Angeles, CA, USA.

## Abstract

**Background**. With the rapid expansion of value-based care in the U.S. health-care industry, addressing health disparities has become a central focus for stakeholders. Social determinants of health (SDoH) indices serve as critical tools for assessing neighborhood inequities and informing programs aimed at managing patient health and costs. The Area Deprivation Index (ADI), recently endorsed by the U.S. Department of Health & Human Services, is widely used in these efforts. However, the ADI's reliance on housing-related variables, such as median home value, raises concerns about its applicability, particularly in regions with high property values. This reliance may obscure inequalities and poor health outcomes in communities with high living costs, limiting its effectiveness for population health initiatives led by accountable care organizations (ACOs).

**Methods**. To address these limitations, we developed the balanced ADI (bADI), a novel SDoH index designed to reduce dependency on housing-related variables by standardizing construction factors. We evaluated the bADI's performance through extensive benchmarking using data from millions of Medicare beneficiaries enrolled in fee-for-service (FFS) and Medicare Advantage (MA) programs. Correlation analyses were conducted to assess the bADI's relationship with clinical outcomes, life expectancy, healthcare utilization, and costs, and its performance was compared to that of the ADI.

**Results**. The bADI demonstrated stronger correlations with clinical outcomes and life expectancy compared to the ADI. Notably, the bADI exhibited reduced dependence on housing variables in high-cost-of-living regions. Benchmarking results revealed that the bADI outperformed the ADI in predicting healthcare utilization and costs. While ADI research suggested that both the least and most disadvantaged neighborhoods incur higher annual healthcare costs, the bADI provided more nuanced insights: the least and most disadvantaged groups experienced higher and lower costs, respectively, relative to other patients.

**Conclusions**. The bADI's robust associations with healthcare outcomes highlight its potential to improve cost management and care management strategies aimed at reducing health disparities. By adopting the bADI, ACO programs could implement budget-neutral redistributions of resources, reallocating funding from less disadvantaged areas to the most deprived regions. This approach promises to enhance equity and effectiveness in addressing social determinants of health within the healthcare system.

## 1. Introduction

Recently, both the health care community and government agencies have begun to increase their focus on policies and programs which address inequities in the healthcare system. As a result, there is a strong need for comprehensive, accurate, and easily accessible measures of SDoH to guide resource allocation (1). In line with these changes, the Centers for Medicare and Medicaid Services (CMS) recently announced a new multi-level equity policy to allocate resources in a cost-neutral manner based on neighborhood disparities measured by SDoH indices (2).

Under this new policy, a specific SDoH index called the ADI helps underserved communities in CMS' two primary ACO programs: Realizing Equity, Access, and Community Health (ACO REACH) and the Medicare Shared Savings Program (MSSP) (3). Specifically, in ACO REACH, financial benchmarks are adjusted based on patients' neighborhood deprivation as measured by ADI. For MSSP, ADI is used to calculate the health equity adjustment (HEA) for the quality performance score of ACOs and the Advance Investment Fund (AIF) for new ACOs (3).

ADI was first introduced in Singh (4) as a weighted linear combination of 17 variables centered on five neighborhood characteristics: education, employment, housing, resource access, and socioeconomic status (SES) (5). This index was one of three small-area deprivation indices approved for policy application by the Office of the Assistant Secretary for Planning and Evaluation (ASPE) with the United States Department of Health and Human Services in late 2022 (6). Furthermore, a recent systematic comparison of small-area disadvantage indices in the United States discovered that of the five (5) SDoH indices (including those approved by ASPE), ADI was the most closely related to a wide range of healthcare outcomes (5).

While the relationship between ADI and clinical outcomes have been extensively studied in various applications, including 30 day rehospitalization (7,8), disease progression (9), chronic pain (10) and mortality (11), research into its association with financial outcomes and resource utilization is limited (12). Understanding the extent to which ADI and other SDOH indices influence such outcomes can assist policymakers, such as CMS, in determining how to best incorporate SDOH indices into financial aid programs (13). The first research gap that this study aims to address is a thorough investigation into the association of ADI and other SDOH indices with financial outcomes.

Despite the numerous advantages of the ADI, its most significant disadvantage is its final value is primarily determined by four housing variables expressed in dollar terms, particularly the median home value. This disadvantage is particularly pronounced in high cost of living areas, where housing costs significantly dominate other factors. For example, previous research found that in New York City, median home value has a 0.98 correlation with ADI (1). New York City contains both the richest and the poorest congressional districts in the country. The South Bronx is one of the poorest congressional districts in the country, ranking in the bottom two national quintiles for life expectancy. In contrast, the Upper East Bronx is one of the most affluent congressional districts in the country, with life expectancy falling in the top percentile nationwide. However, under CMS' MSSP or REACH regulations, both areas would be considered more advantaged due to their ADI values both being the bottom quintiles (3).

While some studies have suggested addressing the aforementioned issue by standardizing the variables (14), to the best of our knowledge, no research has implemented and evaluated this modification to the index. A comprehensive evaluation is essential, given concerns about the effectiveness of such a modification compared to the original version of the index (2). Therefore, the first contribution of this study is the implementation of a modified index, which we refer to as the balanced ADI (bADI), and the second is a thorough evaluation of its performance relative to the original ADI, using real-world patient data from care delivery organizations.

## 2. Background

### 2.1. Area deprivation index

Socioeconomic disadvantage refers to the challenges patients may face in their neighborhoods due to factors such as low income, limited education, and substandard living conditions (15). Assessing an individual's socioeconomic status can be a resource-intensive and often impractical task (16,17). As an alternative, a social determinants of health (SDoH) framework was proposed to evaluate socioeconomic status at the neighborhood level, using criteria such as the concentration of poverty as a proxy (18). Over time, this framework has led to the development of various indices to measure socioeconomic disadvantage at the neighborhood level, including the Social Vulnerability Index (SVI), the Social Deprivation Index (SDI), and the ADI (19). A comprehensive comparison of these three indexes is provided in Table A1.

In the healthcare industry, there is an increasing demand to understand the relationship between SDoH indices and healthcare outcomes to make informed decisions when making policy changes. (20). Recently, the office of the Assistant Secretary for Planning and Evaluation (ASPE) at the U.S. Department of Health & Human Services (HHS) endorsed SDI, SVI, and ADI for policy application (21). In response, CMS announced it would use ADI to help underserved communities in its two primary ACO programs: REACH and MSSP (2). ACO REACH adjusts financial benchmarks based on patients' neighborhoods using the ADI national percentile score, which divides patients into deciles ranked from low to high, with the top decile representing the most in need population. CMS implements a progressive approach to redistribute funds from less deprived geographies to the most deprived geographies, increasing spending benchmarks by $30 for the top decile of disadvantage and decreasing spending benchmarks by $6 in the bottom five deciles (22). Furthermore, CMS incorporates the ADI into the MSSP in two key ways to incentivize ACOs to serve disadvantaged populations. First, the ADI is used to calculate a Health Equity Adjustment (HEA) to ACO quality performance scores—ACOs receive a higher HEA if a larger proportion of their patients reside in areas with high ADI values (i.e., the most socioeconomically deprived areas). Second, CMS uses the ADI to determine funding amounts for the Advance Investment Payment (AIP) opportunity available to new ACOs, with those serving fewer patients in high-ADI areas receiving reduced or no funding (3).

ADI was proposed by Singh (4) to construct a composite census tract based socioeconomic index that could allow the monitoring of population health inequalities. A decade later, Kind et al. (7) adapted this index to the census block group level, which was then made publicly available online, making it more accessible to policymakers, researchers, patients, caregivers, and clinicians (23). Particularly, they calculated ADI scores for each U.S. census block group by adding the 17 socioeconomic variables proposed by Singh (4) and weighting them according to their coefficients. These coefficients were calculated with the principal factor extraction method (24). Despite the numerous advantages of the current version of ADI (5), several studies have identified some of its drawbacks (1,3,25). The most significant disadvantage is that, despite the inclusion of 17 variables in the ADI calculation, the final value is primarily determined by four housing variables expressed in dollar terms, particularly median home value (14). The reason for this is that the 17 variables' units were not standardized before being multiplied by their respective weights, causing the housing variable to be the primary driver of this index. This study aims to improve upon the original ADI by first standardizing these factors in the hopes of mitigating this limitation.

## 2.2. Association of ADI with healthcare clinical and financial outcomes

The relationship between the ADI and clinical healthcare outcomes has been investigated in multiple studies (12), using various types of data with different levels of granularity (26). Singh (4) showed that the relationship between their proposed ADI index and county-level healthcare outcomes extracted from population-level analysis and community estimates (PLACES) data provided by the Centers for Disease Control and Prevention (CDC), which reports 36 clinical prevalence indicators on healthcare measures such as chronic diseases, preventive services, and disabilities. This data have also been used in a recent benchmarking study to compare five different SDoH indices at the tract-level (5). Another tract-level investigation was conducted using the U.S. small-area life expectancy estimates project (USALEEP) data that provides estimation on the life expectancy at birth (27). This is particularly useful because life expectancy is frequently used as a primary indicator of overall health in the literature (27), and recent criticisms of ADI have highlighted how, in a city like New York, the ADI is similar overall, but the life expectancy varies significantly by tract (3).

In addition to aggregated health outcomes, such as those provided by PLACES or USALEEP, patient-level analysis remains the most robust method for evaluating the quality of an SDoH index (2). To this end, researchers initially examined the ability of the ADI to predict 30-day rehospitalization, which was later extended to include additional outcomes such as healthcare costs, mortality, and disease progression (7,8). These studies demonstrated that the ADI is associated with readmission risk, particularly at urban teaching hospitals (8). Other researchers explored the relationship between ADI and kidney disease progression (9), as well as its association with long-term mortality following myocardial infarction (11).

While the relationship between SDoH indices, particularly ADI, and clinical outcomes has received extensive attention, research into its association with financial outcomes and resource utilization is limited (12). A thorough understanding of such relationships is especially important given the rapid expansion of pay-for-performance programs, such as CMS ACO programs (see section 2.1), in the US health care market, where resources are invested in efforts to help vulnerable social conditions and improve the value of healthcare.

To the best of our knowledge, Zhang et al. (28) is the only study that has empirically examined the relationship between ADI and financial outcomes using real-world patient data. Specifically, they analyzed the association between total medical costs and ADI among Medicare FFS beneficiaries residing in New York and New Jersey. Building on this work, our study expands the investigation in two key ways. First, we assess the association between financial outcomes and our proposed bADI, comparing its performance to that of the original ADI. Second, we broaden the scope by including patients from diverse regions across the United States, enrolled in both FFS and MA programs, to provide more comprehensive, nationwide evidence of the association.

## 3. Methods

### 3.1. Social barrier index

To develop bADI we extracted the same 17 variables proposed in Singh (4) and adapted in Kind et al. (7) from the American Community Survey (ACS) data (29), which provides five year estimates of income, education, employment, health insurance coverage, and housing costs and conditions for residents of the United States (30). Detailed description of the extracted variables can be found in Table A2.

The data preprocessing steps of Kind et al. (23) to filter unreliable census block groups and impute missing values have not been publicly documented, a limitation that has been noted in the literature (1,5). To replicate the national percentile and state-level decile ADI scores reported by Kind et al., we explored various imputation methods and successfully reproduced the results using the following preprocessing steps. First, unreliable census block groups were removed based on the filtering criteria detailed in Table A3. Second, missing values were imputed using a geographically nested k-Nearest Neighbors approach (31). Specifically, missing variables were imputed using the k nearest block groups within the same census tract, when available; otherwise, the k nearest neighbors within the same county were used. After testing different values of k, we found that k=5 yielded results identical to those reported in Kind et al. (23).

To generate bADI scores, we first standardize the 17 variables using Z-scores. Coefficients are then derived using the principal factor extraction method, in which the inverse of the correlation matrix is multiplied by the factor loadings. This approach is adapted from Singh (4), which provided the original ADI coefficients. Consistent with Singh's methodology, the resulting index is standardized by setting the mean to 100 and the standard deviation to 20. Neighborhoods are then ranked into percentiles based on increasing bADI values.

### 3.2. Association of bADI with housing value

The first set of experiments was carried out to ensure that bADI resolved the main issue with ADI, i.e. its strong dependency with housing values (1) (see section 2.1). Aside from comparing the Pearson correlation between ADI and bADI with housing value for all census counties, we also looked at the correlation for the top 20 metropolitan areas (MPAs) with the most expensive housing values in the United States.

### 3.3. Association of bADI with healthcare clinical outcomes

The second set of experiments compared bADI to ADI in terms of correlation to healthcare clinical outcomes. Since HHS has also approved SVI and SDI for policy application (see section 2.1), they were included in the benchmarking experiments as well. For conducting the comparisons, the overall Pearson correlation between the PLACES' 36 county level clinical outcomes as well as USALEEP's life expectancy and county level values of the benchmarking SDoH indices including bADI, ADI, SVI, SDI were calculated. Aligned with previous studies, (4,5) county-level results were calculated by averaging across all block groups within a county. The same set of experiments were conducted at tract-level using PLACES' tract level 36 clinical outcomes as well as USALEEP's life expectancy.

### 3.4. Association of bADI with healthcare financial outcomes

The third set of experiments evaluated bADI against ADI at the patient-level with respect to financial outcomes. To do this, 100 million medical claims from one million Medicare beneficiaries in the FFS and MA programs from January 2022 to January 2023 in nine different states in the United States were utilized to assess the indexes' relationship with medical expenses and emergency room (ER) visits. This data were collected from the Optum Labs Data Warehouse (OLDW) (32), which contains information from various care delivery organizations (CDOs) in different states. Table 1 shows the descriptive statistics of the included patients.

Table 1. Descriptive statics of Medicare patients included in the healthcare financial outcomes experiments.

| MA Program | State | Total patients (1,000) | Male (%) | Average Age | MA Program | State | Total patients (1,000) | Male (%) | Average Age |
|---|---|---|---|---|---|---|---|---|---|
| FFS | OH | 45 | 44 | 72 | MA | CA | 97 | 48 | 69 |
|  | MO | 36 | 46 | 69 |  | TX | 174 | 47 | 73 |
|  | CA | 108 | 45 | 73 |  | AZ | 39 | 39 | 68 |
|  | TX | 183 | 39 | 74 |  | NV | 41 | 44 | 74 |
|  | AZ | 45 | 41 | 68 |  | CA | 77 | 46 | 67 |
|  | NV | 47 | 45 | 71 |  | OH | 36 | 44 | 66 |
|  | FL | 96 | 48 | 70 |  | CT | 44 | 38 | 73 |
|  | CT | 38 | 45 | 68 |  | MO | 41 | 46 | 72 |

Following Zhang et al. (28), we rescaled the national percentile ranks of both ADI and bADI for census block groups into quintiles. A generalized linear model with a log link and gamma distribution was then employed to compare neighborhoods with the least disadvantaged conditions (quintile 1) and the most disadvantaged conditions (quintile 5) to those with intermediate conditions (quintile 3), which served as the reference group. Consistent with (28), we adjusted for patient characteristics, including age, sex, race, total number of chronic conditions, total number of conditions per CMS Hierarchical Condition Categories (HCC), and the CMS HCC score, which reflects patients' overall health status.

# 4. Results

## 4.1 Association of bADI with housing value

Figure 1 and Table 2 examine the association of ADI and bADI with house value across over 3,200 census counties. As seen, aligned with the literature (1,3), ADI places a heavy emphasis on housing correlation, but bADI mitigates this issue. The raw Pearson correlations of ADI and bADI to the housing value across all US census counties can be found in Figure A1.

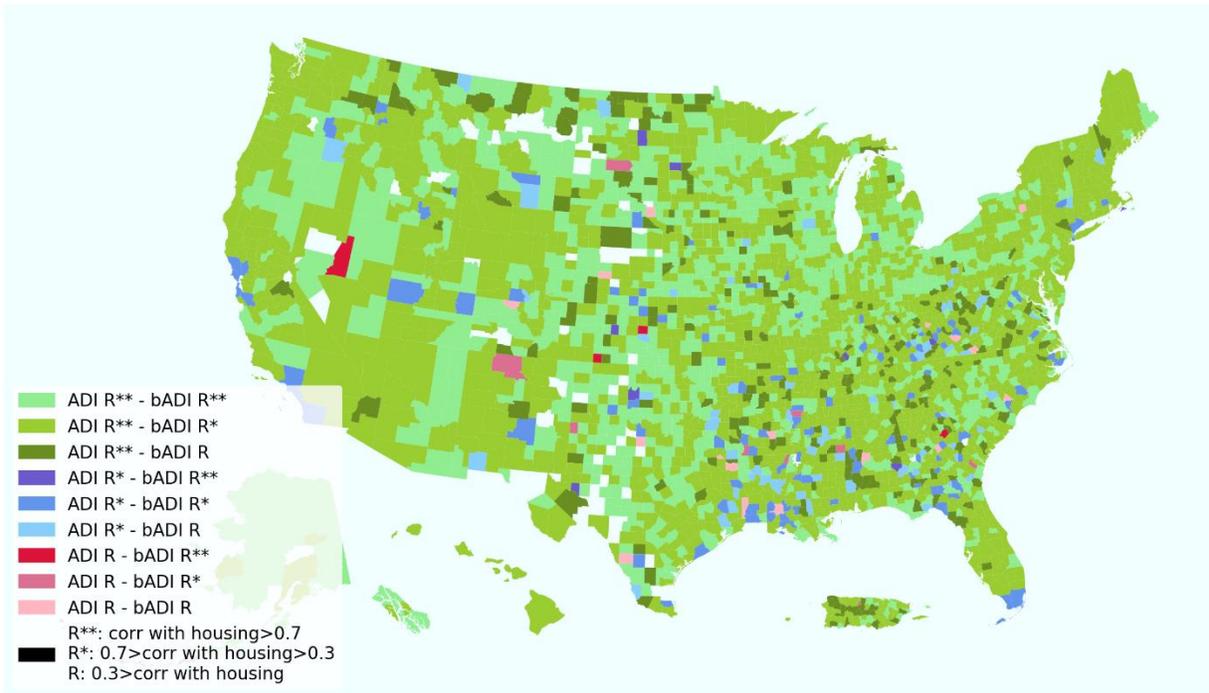

Figure 1. ADI versus bADI correlation comparison per county.

Table 2. Quantile values of correlation of ADI and bADI with housing over all census counties.

|      | Min  | Q1   | Q2   | Q3   | Max  |
| ---- | ---- | ---- | ---- | ---- | ---- |
| **ADI**  | 0.12 | 0.84 | 0.90 | 0.96 | 0.99 |
| **bADI** | 0.10 | 0.46 | 0.58 | 0.69 | 0.98 |

While most concerns have been made about the New York MPA (1,3), Table 3 demonstrates that the ADI housing emphasis issue applies to the majority of the costly MPAs, which is mitigated by bADI. This improvement is achieved by standardizing the 17 constructing variables used to construct the index.

Table 3. Correlation of ADI and bADI with housing for the top 20 most expensive MAPs.

| Metropolitan area | bADI | ADI | Metropolitan area | bADI | ADI |
| --- | --- | --- | --- | --- | --- |
| Barnstable Town, MA | 0.46 | 0.82 | San Diego-Chula Vista-Carlsbad, CA | 0.57 | 0.66 |
| Boston-Cambridge-Newton, MA-NH | 0.42 | 0.74 | San Francisco-Oakland-Fremont, CA | 0.54 | 0.69 |
| Boulder, CO | 0.58 | 0.75 | San Jose-Sunnyvale-Santa Clara, CA | 0.62 | 0.62 |
| Bozeman, MT | 0.59 | 0.88 | San Luis Obispo-Paso Robles, CA | 0.40 | 0.80 |

| | | | | | | |
|---|---|---|---|---|---|---|
| Kahului-Wailuku, HI | 0.54 | 0.74 | Santa Cruz-Watsonville, CA | 0.59 | 0.80 |
| Los Angeles-Long Beach-Anaheim, CA | 0.59 | 0.68 | Santa Maria-Santa Barbara, CA | 0.63 | 0.75 |
| Napa, CA | 0.48 | 0.65 | Santa Rosa-Petaluma, CA | 0.59 | 0.65 |
| New York-Newark-Jersey City, NY-NJ | 0.27 | 0.63 | Seattle-Tacoma-Bellevue, WA | 0.60 | 0.77 |
| Oxnard-Thousand Oaks-Ventura, CA | 0.62 | 0.74 | Urban Honolulu, HI | 0.53 | 0.73 |
| Salinas, CA | 0.70 | 0.74 | Washington-Arlington-Alexandria, DC-VA-MD-WV | 0.56 | 0.80 |

## 4.2. Association of bADI with healthcare clinical outcomes

Table 4 compares the correlation between bADI and 36 clinical outcomes, as well as life expectancy versus ADI, SDI, and SVI. Except for heart disease, cancer, and annual checkups, bADI shows a stronger correlation with outcomes. This is the first evidence that, while bADI was able to resolve the housing emphasis issue (see section 4.1), it allays concerns that a modified version of ADI (after altering the original inputs) might not retain its strong correlation with healthcare outcomes (2). Furthermore, given that life expectancy is frequently referred to as a primary indicator of overall health status in the literature (3,27), bADI provides a significant advantage in this regard as well.

Table A4 provides a more detailed comparison of clinical outcomes by benchmarking six key outcomes across the ten most populated MAPs. The findings are consistent with the county-level analysis presented in Table 4. Additionally, since ADI values in CMS programs are converted into decile buckets for policy implementation (see Section 2.1), the same approach was applied to convert all SDoH indices and clinical outcomes into decile buckets, followed by calculation of the Spearman rank correlation (33) between the SDoH indices and clinical outcomes. As shown in Table A5, the results align with those presented in Table 4.

Table 4. Correlation of bADI, ADI, SDI, and SVI with 36 clinical outcomes and life expectancy.

| Category | Measure | bADI | ADI | SDI | SVI | Category | Measure | bADI | ADI | SDI | SVI |
|---|---|---|---|---|---|---|---|---|---|---|---|
| **Disability** | Any | **0.85** | 0.76 | 0.66 | 0.40 | **Risk Behaviors** | Binge Drinking | **-0.48** | -0.24 | -0.22 | -0.41 |
| | Cognitive | **0.85** | 0.65 | 0.76 | 0.39 | | Smoking | **0.83** | 0.75 | 0.59 | 0.22 |
| | Hearing | **0.67** | 0.57 | 0.48 | 0.17 | | Physical Inactivity | **0.84** | 0.76 | 0.69 | 0.30 |
| | Living | **0.88** | 0.76 | 0.81 | 0.39 | | Sleep <7 hours | **0.63** | 0.57 | 0.53 | 0.36 |
| | Mobility | **0.86** | 0.77 | 0.70 | 0.41 | **Health Status** | General | **0.86** | 0.63 | 0.72 | 0.40 |
| | Self-care | **0.86** | 0.72 | 0.62 | 0.35 | | Mental | **0.76** | 0.67 | 0.57 | 0.36 |
| | Vision | **0.82** | 0.57 | 0.72 | 0.36 | | Physical | **0.87** | 0.64 | 0.79 | 0.38 |
| **Health Outcome** | Teeth Lost | **0.86** | 0.74 | 0.70 | 0.34 | **Prevention** | Annual Checkup | 0.27 | **0.34** | 0.22 | 0.20 |
| | Arthritis | **0.64** | 0.57 | 0.47 | 0.26 | | Cervical Cancer Screening | **-0.52** | -0.42 | -0.23 | -0.11 |
| | COPD | **0.81** | 0.69 | 0.64 | 0.30 | | Cholesterol Screening | **-0.25** | -0.19 | -0.01 | 0.20 |

| | | | | | | | | | |
|---|---|---|---|---|---|---|---|---|---|
| Cancer | -0.25 | -0.04 | **-0.47** | -0.23 | | Colorectal Cancer Screening | **-0.39** | -0.32 | -0.31 | -0.06 |
| CKD | **0.82** | 0.66 | 0.53 | 0.36 | | Core preventive for men | **-0.52** | -0.47 | -0.46 | -0.11 |
| Heart Disease | 0.79 | **0.88** | 0.72 | 0.30 | | Core preventive for women | **-0.55** | -0.37 | -0.21 | -0.15 |
| Current Asthma | **0.55** | 0.44 | 0.32 | 0.19 | | Dental Visit | **-0.81** | -0.75 | -0.65 | -0.37 |
| Depression | **0.36** | 0.26 | 0.28 | 0.21 | | Health Insurance | **0.51** | 0.37 | 0.34 | 0.27 |
| Diabetes | **0.80** | 0.59 | 0.71 | 0.39 | | Mamo | **-0.31** | -0.23 | -0.19 | -0.04 |
| High Blood Pressure | **0.75** | 0.63 | 0.50 | 0.42 | | Taking BP Medication | **0.54** | 0.49 | 0.44 | 0.26 |
| High Cholesterol | **0.47** | 0.39 | 0.39 | 0.35 | | | | | | |
| Obesity | **0.71** | 0.65 | 0.47 | 0.18 | **Life Expectancy** | Life Expectancy | **0.64** | 0.51 | 0.58 | 0.31 |
| Stroke | **0.84** | 0.62 | 0.79 | 0.35 | | | | | | |

COPD: Chronic Obstructive Pulmonary Disease; Mamo: Mammography; CKD: Chronic Kidney Disease.

### 4.3. Association of bADI with healthcare financial outcomes

Figure 2 depicts the cost differential between patients in the least and most advantaged quintiles in terms of total medical expenses for bADI and ADI, from which two inferences may be drawn for FFS and MS patients. First, regardless of region, aligned with the findings of Zhang et al. (28) for New York and New Jersey, using ADI, the least disadvantaged neighborhoods (quintile 1) are associated with higher total costs for FFS patients, whereas the most disadvantaged neighborhoods (quintile 5) have similar total costs to the reference group (quantile 3). However, using bADI, both the least and most disadvantaged areas of FFS patients are associated with greater and lower Medicare costs, respectively, when compared to the reference group. Second, while there is no specific implication for MA patients utilizing ADI across all geographic regions, bADI delivers a comparable implication to FFS patients about total expenses when comparing the most disadvantaged and least advantaged neighborhoods with the reference group. Third, even with bADI, the difference between the most disadvantaged neighborhoods and the reference group is greater than that of the least disadvantaged neighborhoods.

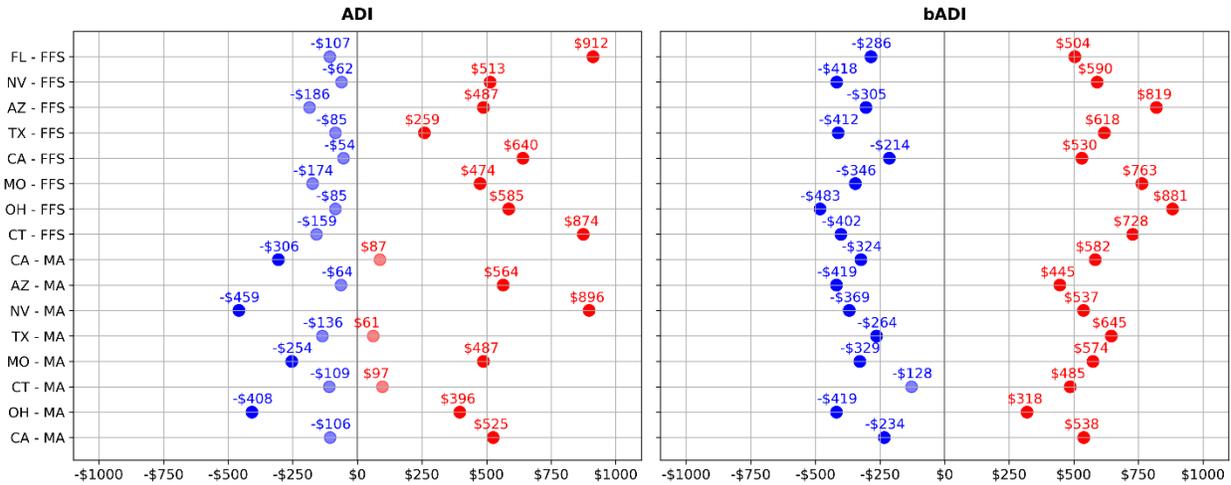

Figure 2. Adjusted associations between ADI and bADI quintiles and total costs between Jan 2022 and Jan 2023. Numbers in the figure indicated the increased or decreased costs associated with quintiles 1 and 5, as compared with quintile 3 (reference group). Dark red or dark blue indicates that the matching p-value was less than 0.05, but light red or light blue indicate otherwise.

Figure 3 shows the percentage difference between patients in the least and most advantaged quintiles in terms of total number of ER visits for bADI and ADI. In terms of bADI we can observe almost similar observation as total cost, where both the least and most disadvantaged areas of FFS and MA patients were associated with greater and lower percent of ER visits, respectively, when compared to the reference group. However, no specific distinctive trend can be observed for ADI. Moreover, same as patients' costs, the difference between the most disadvantaged neighborhoods and the reference group is greater than that of the least disadvantaged neighborhoods using bADI.

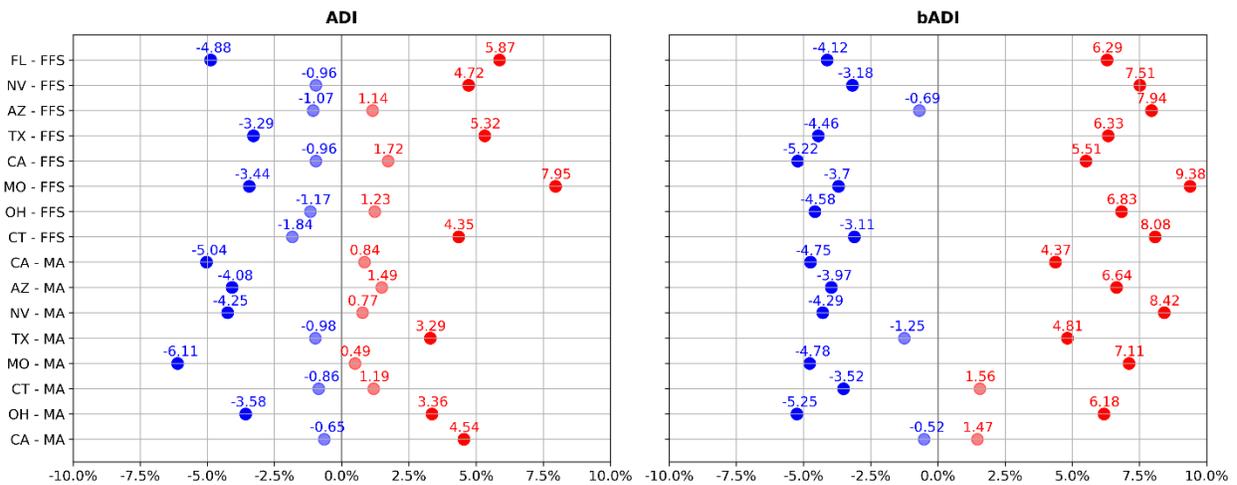

Figure 3. Adjusted associations between ADI and bADI quintiles and number of ER visits between Jan 2022 and Jan 2023. Numbers in the figure indicated the percent increased or decreased number of ER visits associated with quintiles 1 and 5, as compared with quintile 3 (reference group). Dark red or dark blue indicates that the matching p-value was less than 0.05, but light red or light blue indicate otherwise.

# 5. Discussion

The research had three primary objectives. First, we addressed the ADI index's strong dependency with house value by standardizing the construction factors. Second, we performed an extensive evaluation to ensure that the new bADI index demonstrates significant association with previously observed clinical outcomes in the literature. Third, for the first time, we include evaluations to establish the relationship between SDoH indices, notably ADI and bADI, and patient expenses.

The initial set of experiments demonstrate that bADI has a much lower correlation with housing value in all U.S. census counties than ADI. This finding was supported by more detailed evaluations considering the most expensive MAPs. These results demonstrate that bADI provides more balanced attention to the different input categories: education, employment, housing, resource access, and income.

The second set of experiments show that bADI improves upon ADI's high association with clinical outcomes. This was demonstrated using county-level data as well as considering the most populous MAPs. This suggests that balanced attention to the various dimensions of a patient's social status (rather than just housing value) improves the association between an SDoH index and clinical outcomes.

The third set of evaluations confirmed a strong association between the bADI and financial outcomes. This was demonstrated through a comprehensive series of experiments involving both FFS and MA beneficiaries across multiple states in the United States. These findings support the policy application of SDoH indices in population health initiatives, such as CMS's ACO programs aimed at addressing healthcare disparities through budget-neutral resource redistribution. Moreover, the observed correlation between higher bADI scores—indicating greater socioeconomic disadvantage—and increased healthcare spending may suggest that disadvantaged individuals face barriers to accessing care and often have unmet needs for essential health services. By identifying the specific social factors contributing to limited access, policymakers and healthcare leaders can more effectively address the unmet healthcare needs of these populations.

This research has several limitations. First, in developing the bADI, we did not incorporate any new variables beyond those included in the original ADI. Future work should consider integrating additional factors—such as food access, crime rates, and transit availability—by leveraging the more than 150 social determinants identified in the literature (27). Second, our clinical outcome evaluations were limited to county- or MAP-level analyses. While these provide useful insights into the association between bADI and healthcare outcomes, patient-level analyses would offer a more granular and precise understanding. Third, although this study examined associations between bADI and various healthcare outcomes, it did not attempt to establish causal relationships. Identifying causal links could help uncover specific factors that contribute to healthcare disparities.

# 6. Conclusion

This study aimed to develop an SDoH index that places balanced emphasis on education, employment, housing, resource access, and income. The findings indicate that this enhanced ADI index is strongly associated with patients' clinical outcomes. Additionally, we provide detailed national evidence demonstrating a strong relationship between the index and healthcare expenditures.

# 7. Author Statements

**Ethical approval**. Ethical approval was not necessary for this study since it did not involve human or animal subjects or any unanonymized data. The OLDW is a real-world database that contains only de-identified information.

**Availability of data and materials**. The ACS datasets generated bADI during the current study are available in the census.gov repository (29). Also, OLDW datasets analyzed during the current study are available in the Optum Labs repository (32).

# Appendix

Table A1. Comparison of ADI, SVI and SDI.

| Feature / Index | ADI | SVI | SDI |
|---|---|---|---|
| **Developer** | University of Wisconsin – Health Innovation Program | Centers for Disease Control and Prevention | Robert Graham Center |
| **Purpose** | Measure neighborhood-level disadvantage | Disaster preparedness, public health risk | Measure socioeconomic disadvantage |
| **Level of Geography** | Census block group / ZIP code | Census tract | ZIP code / Census tract |
| **Total Variables** | 17 | 15 | 7 |
| **Data Source** | ACS / U.S. Census | ACS | ACS / U.S. Census |
| **Scoring Method** | Weighted composite index + national percentile | Percentile ranking by theme + overall | Composite score from standardized indicators |
| **Focus** | Socioeconomic deprivation and health outcomes | Vulnerability to public health threats | Access to resources and social disadvantage |
| **Variables** | **Income**: Percent below poverty level, Median family income, Income disparity | **Socioeconomic Status**: Percent below poverty, Per capita income, Percent with no high school diploma, Percent unemployed | **Income**: Percent living in poverty |
| | **Education**: Percent with < 9 years of education, Percent with high school diploma only, Percent with ≥ 12 years of education | **Household composition and Disability**: Percent aged 65 and older, Percent aged 17 and younger, Percent with a disability, Percent of single-parent households (with children under 18) | **Education**: Percent with less than 12 years of education |
| | **Employment**: Percent unemployed, Percent of adults not in labor force | **Minority Status & Language**: Percent of minority (non-white) population, Percent who speak English 'less than well' | **Employment**: Percent non-employed adults under 65 |
| | **Housing Quality**: Percent of households with no vehicle, Percent of households with no telephone, Percent of households without complete plumbing facilities, Percent of occupied housing units without a kitchen, Percent of renter-occupied housing units, Percent of housing units that are mobile homes, Median home value, Median gross rent, Percent of housing units with crowding (>1 person per room) | **Housing type & Transportation**: Percent of housing in multi-unit structures, Percent of mobile homes, Percent of households with more people than rooms, Percent of households with no vehicle, Percent of individuals in group quarters | **Housing Quality**: Percent single-parent households with dependents under 18, Percent living in a rented housing unit, Percent living in overcrowded housing, Percent of households without a car |

Table A2. Constructing variables of bADI.

| Category | Concept | ACS 5-Year variable group |
|---|---|---|
| SES | Median family income ($) | B19113 |
| | Income disparity (%) | B19001 |
| | Families below poverty level (%) | B17010 |
| | Population below 150% of poverty level (%) | C17002 |
| | Single parent households with dependents under 18 (%) | B09002 |
| Resource access | Households without a motor vehicle (%) | B25044 |
| | Households without a telephone (%) | B25043 |
| | Occupied housing units without complete plumbing (%) | B25016 |
| Housing | Owner occupied housing units (%) | B25003 |
| | Households with more than 1 person per room (%) | B25014 |
| | Median monthly mortgage ($) | B25088 |
| | Median gross rent ($) | B25064 |
| | Median home value ($) | B25077 |
| Employment | Employed person 16 or older in white collar jobs (%) | C24010 |
| | Civilian labor force unemployed (aged 16 or over) (%) | B23025 |
| Education | Population aged 25 or older with no high school (%) | B15003 |
| | Population aged 25 or older with at least a high school education (%) | B15003 |

Table A3. Filtering variables of census block groups for creating bADI.

| Filter | ACS 5-Year variable group |
|---|---|
| Fewer than 100 persons | B01003 |
| Fewer than 30 housing units | B25001 |
| Greater than 33% of the population living in group quarters | B09019 |

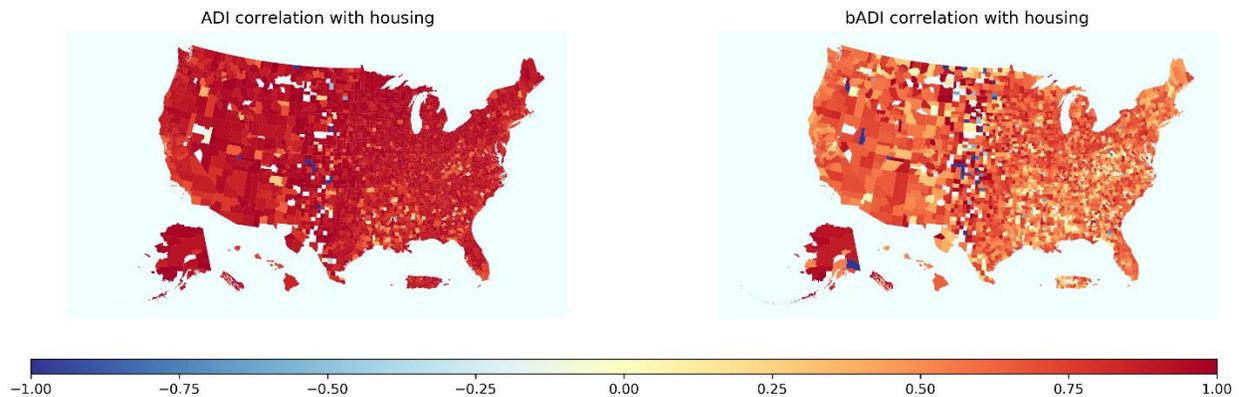

Figure A1. Pearson correlation between ADI and housing versus bADI and housing.

Table A4. Correlation of bADI, ADI, SDI, and SVI with six selected clinical outcomes.

| Measure | Metro | bADI | ADI | SDI | SVI | Measure | bADI | ADI | SDI | SVI |
|---|---|---|---|---|---|---|---|---|---|---|
| **Heart Disease** | Atlanta-Sandy Springs-Roswell, GA | **0.58** | 0.54 | 0.46 | 0.49 | **Health Insurance** | **0.86** | 0.75 | 0.76 | 0.74 |
| | Boston-Cambridge-Newton, MA-NH | 0.40 | **0.48** | 0.20 | 0.39 | | **0.88** | 0.61 | 0.79 | 0.81 |
| | Chicago-Naperville-Elgin, IL-IN | **0.59** | 0.52 | 0.37 | 0.49 | | **0.76** | 0.58 | 0.44 | 0.42 |
| | Dallas-Fort Worth-Arlington, TX | 0.49 | **0.53** | 0.45 | 0.51 | | **0.88** | 0.79 | 0.83 | 0.79 |
| | Houston-Pasadena-The Woodlands, TX | **0.66** | 0.56 | 0.56 | 0.55 | | **0.87** | 0.77 | 0.74 | 0.69 |
| | Los Angeles-Long Beach-Anaheim, CA | **0.29** | 0.17 | 0.15 | 0.22 | | **0.90** | 0.61 | 0.80 | 0.77 |
| | New York-Newark-Jersey City, NY-NJ | **0.31** | 0.27 | 0.17 | 0.25 | | **0.79** | 0.49 | 0.70 | 0.71 |
| | Philadelphia-Camden, PA-NJ-DE-MD | 0.43 | **0.48** | 0.36 | 0.46 | | **0.80** | 0.77 | 0.77 | 0.76 |
| | Phoenix-Mesa-Chandler, AZ | 0.29 | **0.42** | 0.20 | 0.26 | | **0.89** | 0.72 | 0.73 | 0.79 |
| | Riverside-San Bernardino-Ontario, CA | 0.26 | **0.49** | 0.20 | 0.18 | | **0.81** | 0.49 | 0.77 | 0.73 |
| **Asthma** | Atlanta-Sandy Springs-Roswell, GA | **0.88** | 0.71 | 0.72 | 0.80 | **Mental Health** | **0.90** | 0.82 | 0.87 | 0.83 |
| | Boston-Cambridge-Newton, MA-NH | **0.82** | 0.65 | 0.61 | 0.62 | | **0.90** | 0.70 | 0.72 | 0.68 |
| | Chicago-Naperville-Elgin, IL-IN | **0.78** | 0.71 | 0.68 | 0.63 | | **0.87** | 0.78 | 0.71 | 0.73 |
| | Dallas-Fort Worth-Arlington, TX | **0.84** | 0.72 | 0.71 | 0.79 | | **0.92** | 0.86 | 0.86 | 0.84 |
| | Houston-Pasadena-The Woodlands, TX | **0.80** | 0.69 | 0.67 | 0.75 | | **0.92** | 0.85 | 0.89 | 0.85 |
| | Los Angeles-Long Beach-Anaheim, CA | **0.71** | 0.56 | 0.65 | 0.59 | | **0.88** | 0.65 | 0.83 | 0.76 |
| | New York-Newark-Jersey City, NY-NJ | **0.69** | 0.27 | 0.63 | 0.60 | | **0.82** | 0.30 | 0.77 | 0.71 |
| | Philadelphia-Camden, PA-NJ-DE-MD | **0.89** | 0.85 | 0.73 | 0.77 | | **0.93** | 0.85 | 0.67 | 0.69 |
| | Phoenix-Mesa-Chandler, AZ | **0.85** | 0.73 | 0.65 | 0.61 | | **0.93** | 0.68 | 0.35 | 0.39 |
| | Riverside-San Bernardino-Ontario, CA | **0.81** | 0.64 | 0.60 | 0.61 | | **0.92** | 0.56 | 0.82 | 0.74 |
| **Diabetes** | Atlanta-Sandy Springs-Roswell, GA | **0.80** | 0.75 | 0.73 | 0.74 | **Physical Inactivity** | **0.93** | 0.85 | 0.66 | 0.66 |
| | Boston-Cambridge-Newton, MA-NH | **0.76** | 0.57 | 0.61 | 0.63 | | **0.93** | 0.70 | 0.60 | 0.67 |
| | Chicago-Naperville-Elgin, IL-IN | **0.80** | 0.72 | 0.69 | 0.62 | | **0.92** | 0.81 | 0.84 | 0.84 |
| | Dallas-Fort Worth-Arlington, TX | **0.86** | 0.76 | 0.70 | 0.71 | | **0.92** | 0.87 | 0.87 | 0.85 |
| | Houston-Pasadena-The Woodlands, TX | **0.87** | 0.75 | 0.75 | 0.72 | | **0.93** | 0.87 | 0.88 | 0.86 |
| | Los Angeles-Long Beach-Anaheim, CA | **0.85** | 0.52 | 0.68 | 0.71 | | **0.93** | 0.64 | 0.84 | 0.84 |
| | New York-Newark-Jersey City, NY-NJ | **0.77** | 0.65 | 0.57 | 0.52 | | **0.87** | 0.69 | 0.78 | 0.81 |
| | Philadelphia-Camden, PA-NJ-DE-MD | **0.88** | 0.77 | 0.71 | 0.73 | | **0.91** | 0.89 | 0.84 | 0.84 |
| | Phoenix-Mesa-Chandler, AZ | **0.67** | 0.70 | 0.59 | 0.62 | | **0.89** | 0.82 | 0.83 | 0.84 |
| | Riverside-San Bernardino-Ontario, CA | **0.74** | 0.68 | 0.58 | 0.54 | | **0.88** | 0.68 | 0.81 | 0.78 |

Table A5. Spearman correlation between bADI, ADI, SDI, and SVI deciles and the deciles of 36 clinical outcomes and life expectancy.

| Category | Measure | bADI | ADI | SDI | SVI | Category | Measure | bADI | ADI | SDI | SVI |
|---|---|---|---|---|---|---|---|---|---|---|---|
| Disability | Any | **0.84** | 0.66 | 0.76 | 0.41 | Risk Behaviors | Binge Drinking | **-0.75** | -0.31 | -0.64 | -0.41 |
| | Cognitive | **0.86** | 0.65 | 0.77 | 0.41 | | Smoking | **0.82** | 0.74 | 0.58 | 0.24 |
| | Hearing | **0.65** | 0.57 | 0.46 | 0.17 | | Physical Inactivity | **0.83** | 0.76 | 0.69 | 0.30 |
| | Living | **0.90** | 0.68 | 0.83 | 0.41 | | Sleep <7 hours | **0.67** | 0.45 | 0.63 | 0.36 |
| | Mobility | **0.86** | 0.68 | 0.81 | 0.43 | Health Status | General | **0.87** | 0.64 | 0.83 | 0.42 |
| | Self-care | **0.91** | 0.67 | 0.87 | 0.38 | | Mental | **0.76** | 0.55 | 0.68 | 0.37 |
| | Vision | **0.88** | 0.62 | 0.87 | 0.41 | | Physical | **0.87** | 0.64 | 0.80 | 0.40 |
| Health Outcome | Teeth Lost | **0.88** | 0.71 | 0.75 | 0.35 | Prevention | Annual Checkup | 0.26 | 0.24 | **0.27** | 0.25 |
| | Arthritis | **0.61** | 0.51 | 0.49 | 0.27 | | Cervical Cancer Screening | **-0.52** | -0.43 | -0.42 | -0.12 |
| | COPD | **0.83** | 0.69 | 0.66 | 0.31 | | Cholesterol Screening | **-0.32** | -0.29 | 0.03 | 0.24 |
| | Cancer | -0.26 | -0.11 | **-0.44** | -0.24 | | Colorectal Cancer Screening | **-0.39** | -0.35 | -0.28 | -0.06 |
| | CKD | 0.85 | 0.58 | **0.87** | 0.39 | | Core preventive for men | **-0.50** | -0.39 | -0.44 | -0.13 |
| | Heart Disease | **0.90** | 0.74 | 0.73 | 0.33 | | Core preventive for women | **-0.55** | -0.41 | -0.50 | -0.16 |
| | Current Asthma | **0.52** | 0.32 | 0.5 | 0.19 | | Dental Visit | **-0.81** | -0.65 | -0.74 | -0.39 |
| | Depression | **0.32** | 0.18 | 0.26 | 0.20 | | Health Insurance | **0.68** | 0.43 | 0.61 | 0.34 |
| | Diabetes | 0.81 | 0.61 | **0.83** | 0.41 | | Mamo | **-0.31** | -0.25 | -0.19 | -0.04 |
| | High Blood Pressure | **0.74** | 0.62 | 0.69 | 0.42 | | Taking BP Medication | **0.58** | 0.54 | 0.47 | 0.26 |
| | High Cholesterol | **0.52** | 0.40 | 0.49 | 0.35 | Life Expectancy | Life Expectancy | **0.67** | 0.5 | 0.6 | 0.33 |
| | Obesity | **0.66** | 0.68 | 0.49 | 0.18 | | | | | | |
| | Stroke | **0.90** | 0.67 | 0.84 | 0.38 | | | | | | |

COPD: Chronic Obstructive Pulmonary Disease; Mamo: Mammography; CKD: Chronic Kidney Disease.